\pdfoutput=1
\documentclass[aps,groupaddress,superscriptaddress,reprint,amsmath,amssymb,graphicx]{revtex4-2}
\usepackage{graphicx}
\usepackage{physics}
\usepackage{bm}
\usepackage{bbold}
\usepackage{color}
\usepackage{amsthm}

\theoremstyle{remark}

\usepackage{setspace}
\usepackage[capitalize]{cleveref}

\newcommand{\der}[2]{\frac{\mathrm{d} #1}{\mathrm{d}#2}}

\begin{document}
\preprint{APS/123-QED}

\title{Spatially disordered environments stabilize competitive metacommunities}

\author{Prajwal Padmanabha}
\thanks{P.P., G.N., and D.B. contributed equally to this work, and are listed in ascending order of seniority.}
\affiliation{Department of Physics and Astronomy ``Galileo Galilei'', University of Padova, Padova, Italy}
\affiliation{Department of Fundamental Microbiology, University of Lausanne, Lausanne, Switzerland}
\author{Giorgio Nicoletti}
\thanks{P.P., G.N., and D.B. contributed equally to this work, and are listed in ascending order of seniority.}
\affiliation{Laboratory of Ecohydrology, School of Architecture, Civil and Environmental Engineering, École Polytechnique Fédérale de Lausanne, Lausanne 1015, Switzerland}
\author{Davide Bernardi}
\thanks{P.P., G.N., and D.B. contributed equally to this work, and are listed in ascending order of seniority.}
\affiliation{Department of Physics and Astronomy ``Galileo Galilei'', University of Padova, Padova, Italy}
\affiliation{Istituto Nazionale di Fisica Nucleare, Sezione di Padova, Padova 35131, Italy}
\affiliation{National Biodiversity Future Center, Palermo 90133, Italy}
\author{Samir Suweis}
\affiliation{Department of Physics and Astronomy ``Galileo Galilei'', University of Padova, Padova, Italy}
\affiliation{Istituto Nazionale di Fisica Nucleare, Sezione di Padova, Padova 35131, Italy}

\author{Sandro Azaele}
\affiliation{Department of Physics and Astronomy ``Galileo Galilei'', University of Padova, Padova, Italy}
\affiliation{Istituto Nazionale di Fisica Nucleare, Sezione di Padova, Padova 35131, Italy}
\affiliation{National Biodiversity Future Center, Palermo 90133, Italy}

\author{Andrea Rinaldo}
\affiliation{Laboratory of Ecohydrology, School of Architecture, Civil and Environmental Engineering, École Polytechnique Fédérale de Lausanne, Lausanne 1015, Switzerland}
\affiliation{Department of Civil, Environmental and Architectural Engineering, University of Padova, Padova 35131, Italy}

\author{Amos Maritan}
\affiliation{Department of Physics and Astronomy ``Galileo Galilei'', University of Padova, Padova, Italy}
\affiliation{Istituto Nazionale di Fisica Nucleare, Sezione di Padova, Padova 35131, Italy}
\affiliation{National Biodiversity Future Center, Palermo 90133, Italy}

\begin{abstract}
\noindent Metapopulation models have been instrumental in demonstrating the ecological impact of landscape structure on the survival of a focal species in complex environments. However, extensions to multiple species with arbitrary dispersal networks often rely on phenomenological assumptions limiting their scope. Here, we develop a multilayer network model of competitive dispersing metacommunities to investigate how spatially structured environments impact species coexistence and ecosystem stability. We show that homogeneous environments always lead to monodominance unless all species' fitness parameters are in an exact trade-off. However, this precise fine-tuning does not guarantee coexistence in generic heterogeneous environments. By introducing general spatial disorder in the model, we solve it exactly in the mean-field limit, finding that stable coexistence becomes possible in the presence of strong disorder. Crucially, coexistence is supported by the spontaneous localization of species through the emergence of ecological niches. Our results remain qualitatively valid in arbitrary dispersal networks, where topological features can improve species coexistence. Finally, we employ our model to study how correlated disorder promotes spatial ecological patterns in realistic terrestrial and riverine landscapes. Our work provides a novel framework to understand how landscape structure enables coexistence in metacommunities by acting as the substrate for ecological interactions.
\end{abstract}

\maketitle

\noindent Predicting the effect of landscape and habitat changes, including fragmentation, on the dynamics of interacting species is a pressing and paramount challenge \cite{ceballos2015accelerated, tilman2017future, isbell2017linking, hanski1998, hanski1999}. However, a comprehensive understanding of the key processes that foster biodiversity of ecosystems in the presence of spatial disturbances remains largely elusive to date \cite{DurrettLevin1994,allesina2012stability,rinaldo2020}. Though several mechanisms for coexistence and maintenance of biodiversity have been proposed \cite{chesson2000mechanisms, loreau2013biodiversity,levin2000multiple}, studies validating them at a local scale vastly outnumber the spatial counterpart \cite{bjorkman2018plant, hillerislambers2012rethinking}. This poses a fundamental limit to our understanding of the composition of ecological communities across spatiotemporal scales and their relation to habitat heterogeneity. Constructing a framework for spatially structured ecosystems is, in general, a formidable and challenging task due to the complexity of species interactions and their role in determining ecosystem stability \cite{allesina2012stability, mougi2012diversity, suweis2014disentangling}, the influence of ever-changing environmental fluctuations shaping population dynamics \cite{chesson1997roles}, and the effects of the landscape structure driving ecological patterns \cite{carrara2012dendritic}. Such challenges are complicated by the simultaneous presence of both short-range dynamics of intra- and interspecific interactions as well as long-range colonization and migration processes. 

In this context, models of metapopulations have proven to be remarkably successful in predicting the survival of a single focal species in complex landscapes of interconnected habitat patches \cite{hanski1998,hanski1999,keymer2000extinction,ova02,rinaldo2020,keymer2000extinction}. In the presence of colonization and extinction events, the seminal work by Hanski and Ovaskainen \cite{HanOva2000} has shown that the long-term survival of a species is quantified by a single landscape measure, named \emph{metapopulation capacity}, related to the landscape features and the underlying dispersal pathways through which the species individuals move \cite{NicPad2023}. Metapopulation capacity subsumes the general viability of a focal species in a given environment, being the leading eigenvalue of a suitable landscape matrix determining the threshold of persistence-free equilibrium \cite{hanski1999}. Such dispersal networks, characterizing the relationships between patches, act as the template for ecological strategies \cite{southwood1977habitat}. They drive a population's dynamics, stability, and persistence in both theoretical \cite{sole2006, ovaskainen2004metapopulation, urban2001landscape, may2019stability,marquet2014theory,gilarranz2012spatial} and field studies \cite{rayfield2023spread,holyoak2000habitat, bevanda2015landscape,staddon2010connectivity,arancibia2022network}.

Yet, how the complex interplay between ecological interactions and landscape structure shapes ecological metacommunities is still an open question \cite{zarnetske2017interplay}. The presence of mutualistic and competitive interactions leads to large-scale fluctuations, both at local and global scales \cite{roy2020complex, pearce2020stabilization,  altieri2022effects}, while niche differences arising due to interspecific tradeoffs act as a stabilizing force to promote coexistence \cite{luo2022multispecies,anceschi2019neutral}. Crucially, through its interplay with different types of interactions, dispersal can benefit coexistence \cite{gravel2016stability} and rescue habitats from extinction \cite{giulia2024interactions}, but also destabilize complex ecosystems \cite{baron2020dispersal}. Although spatial heterogeneity is generally accepted to favor biodiversity, empirical studies have found it may have both positive and negative effects \cite{tamme2010environmental}. These contrasting results partly arise due to the complex relationship between the different mechanisms at play in spatially extended ecosystems. At a local scale, mutualistic interactions are necessary to promote coexistence \cite{kehe2021positive, grilli2017feasibility,gross2008positive,piccardi2019toxicity}. However, in a spatial setting, under what conditions are dispersal and spatial structures beneficial despite competition, or detrimental despite mutualism, remains a fundamental open question.

In this work, we address these shortcomings by developing a general model of ecological metacommunities derived from an underlying individual-level description, where species compete for limited space in multiple habitat patches with varying environments. Patches are connected by a dispersal network leading to global colonization dynamics described by an explicit dispersal kernel. Under these dynamics, we show that a spatially heterogeneous generalization of Hanski and Ovaskainen's metapopulation capacity fails to predict species' survival in a metacommunity, which, even in simple cases, depends on the average fitness of all other species. We analytically prove that, in homogeneous environmental conditions, only the species with the largest fitness survives with coexistence arising only when species are involved in a fine-tuned dispersal-death trade-off. However, we prove that stable coexistence is attainable in sufficiently heterogeneous environments, i.e., in disordered environments where habitat patches are sufficiently different from each other. Crucially, this coexistence stems from the spontaneous emergence of ecological niches, which we quantify by the localization of each species in different habitats. Although our analytical results are rigorously derived in the mean-field limit of large disordered landscapes, we show numerically that they remain good approximations even for smaller ecosystems with varying spatial structures and environments. In particular, we find that structured dispersal networks are often beneficial to coexistence. Furthermore, we show how correlated environmental disorder leads to the formation of spatial ecological patterns in realistic dispersal networks of terrestrial and aquatic landscapes, supporting both coexistence and increased total population. Our findings underscore the complex interplay between landscape heterogeneity and species survival and coexistence, offering new insights into biodiversity preservation in fragmented habitats.

\begin{figure}[t]
	\centering
	\includegraphics[width=\columnwidth]{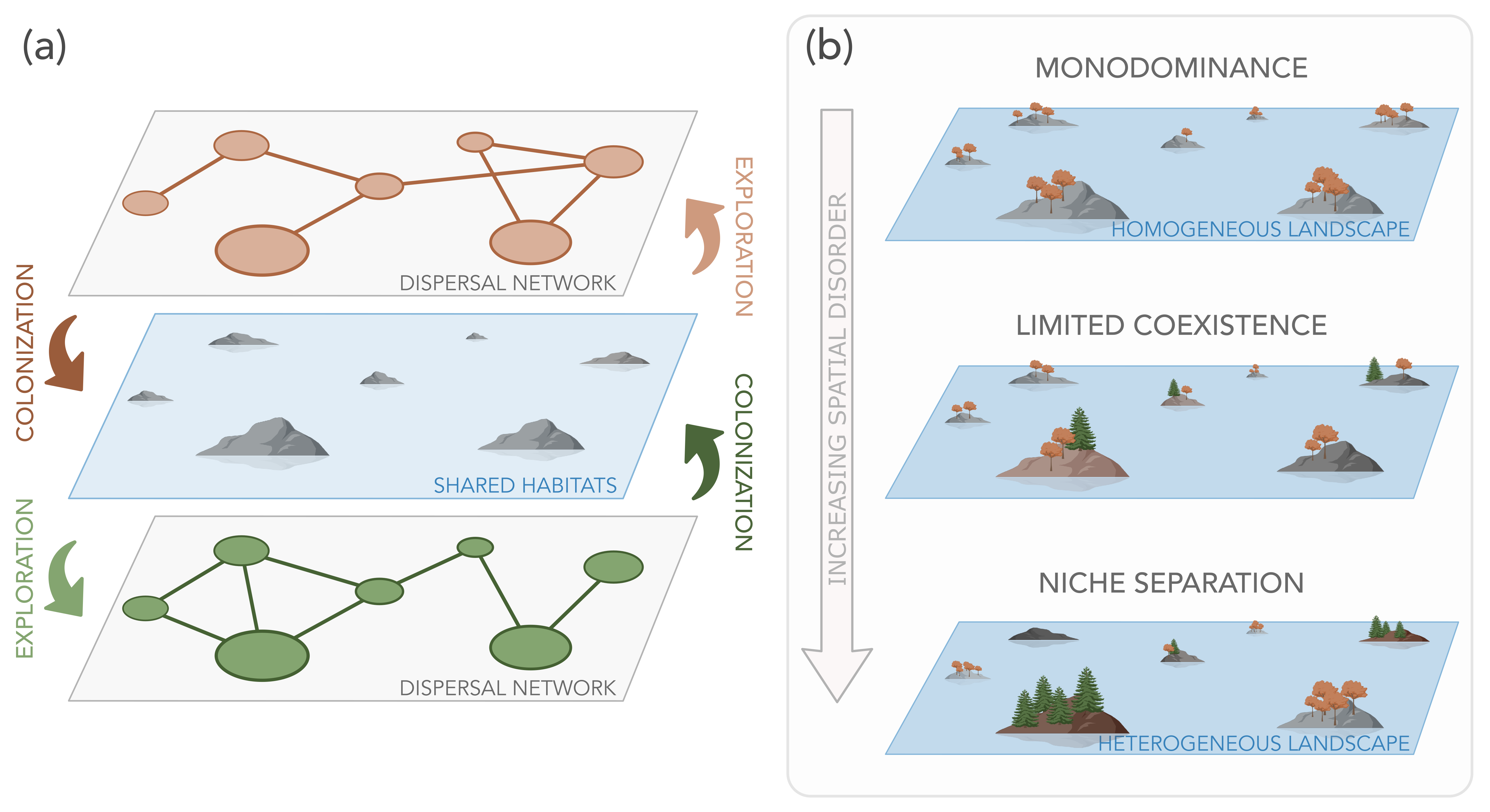}
	\caption{\textbf{In a general model of dispersing metacommunities with competition for space, stable coexistence states emerge in heterogeneous landscapes.} (a) We consider a multilayer dispersal network, where each layer describes the dispersal of a species. Nodes represent shared habitat patches where the species settle and compete for a finite amount of space. (b) In homogeneous landscapes, where habitat patches are equivalent, only the fittest species survives. However, if the quenched disorder modeling landscape heterogeneity is high enough, the coexistence of a large number of species becomes possible through the spontaneous emergence of habitat niches.}
	\label{fig:sketch}
\end{figure}

\section*{Results}
\subsection*{Multilayer network model for dispersing metacommunities}
\noindent We describe the dynamics of $S$ species in $N$ habitat patches, each with a finite number of colonizable sites. Individuals of each species can explore different patches through a shared or species-specific dispersal network. This microscopic description corresponds to a multilayer network dynamics of local and global processes occurring at different scales (see Methods). Assuming that the number of colonizable sites is large and that exploration is fast compared to colonization and death, we explicitly derive the time evolution of the fraction of space occupied by species $\alpha$ in patch $i$, $p_{\alpha i}$ as
\begin{equation}
	\der{p_{\alpha i}}{t} = -e_{\alpha i} p_{\alpha i} + \left(1-\sum_{\beta=1}^{S}p_{\beta i}\right)\sum_{j=1}^{N} K_{\alpha, ij} \, p_{\alpha j},
	\label{eqn:model-general}
\end{equation}
where $K_{\alpha, ij}$ is a species-specific dispersal kernel that describes colonization, and $e_{\alpha i}$ is the local (within patch) extinction rate of species $\alpha$ in patch $i$. Remarkably, we can derive an explicit expression for the kernel that effectively considers all possible paths connecting two patches with an appropriate weight \cite{NicPad2023}. Thus, $K_{\alpha, ij}$ quantifies the rate at which individuals of species $\alpha$ generated in patch $j$ explore the network and eventually colonize patch $i$. The term $(1-\sum_{\beta=1}^{S}p_{\beta i})$ represents the free space in patch $i$, which introduces competition between species.

If only a single focal species were present, the long-time behavior of the system would be determined by a measure called metapopulation capacity \cite{HanOva2000}. For constant extinction rate $e$, the seminal work of Hanski and Ovaskainen \cite{HanOva2000} showed that the metapopulation capacity is the largest eigenvalue $\lambda_M$ of a suitable landscape matrix determining the global extinction threshold for the focal species. Indeed, if $\lambda_M > e$, the species survives; otherwise, it goes extinct in all patches. In heterogeneous landscapes, where $e_i$ depends explicitly on the patches, we prove instead that survival is possible only when $\lambda>1$, where $\lambda$ is the largest eigenvalue of the matrix $\mathbb K \mathbb E^{-1}$, with $E _{ij} =e_i\delta_{ij}$ (see Supplementary Information). Thus, the metapopulation capacity depends on all $e_i$ at once, which underlines the significance of variations in local extinction rates. This suggests that landscape heterogeneity plays a major role in determining the survival and, as we will show, the coexistence of multiple species. We sketch the model and these ideas in \cref{fig:sketch}.

\subsection*{Fine-tuned coexistence in homogeneous landscapes}
\noindent In \cref{eqn:model-general}, landscape heterogeneity enters through both the dispersal pathways determining $K_{\alpha, ij}$ and the patch- and species-dependent extinction rates $e_{\alpha i}$. To disentangle their effects, we first consider the homogeneous case in which all habitat patches have the same extinction rate, i.e., $e_{\alpha i} = e_\alpha$ for all $i$. 

\begin{figure}[t]
	\includegraphics[width=\columnwidth]{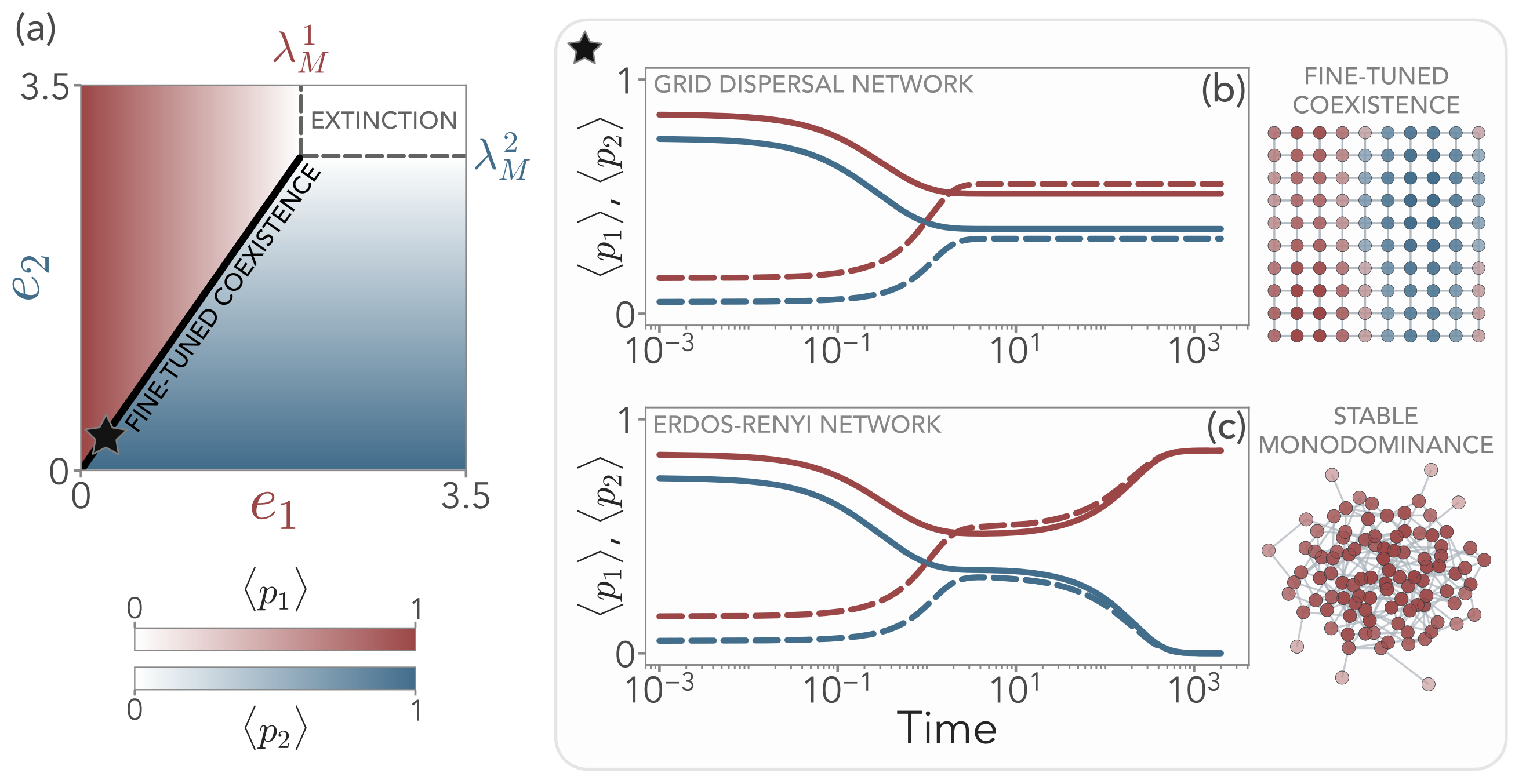}
	\caption{\textbf{With homogeneous extinction rates, a single species dominates in the long-time limit and general stable coexistence is not possible except in a fine-tuned regime.} (a) Results for a grid dispersal network with two species (red and blue) with extinction rates $e_1$ and $e_2$, respectively, equal in all patches. If an extinction rate exceeds the corresponding metapopulation capacity $\lambda_M^\alpha$, a species goes extinct (upper right corner). Coexistence is only possible if the ratio $\ev{K}_\alpha / e_\alpha$ is equal for all species (black line), where the average kernel $\ev{K}_\alpha$ cannot depend on patches due to the translational invariance of the underlying dispersal network. (b) With an equal ratio for all species, the stationary coexistence state depends on the initial conditions (solid and dashed lines) and corresponds to a central manifold. Note that, since exploration is most effective between neighboring patches, the two species survive in separated regions of the dispersal network. (c) In general dispersal networks, the stable state is one in which one species dominates and all others go extinct, independently of the initial conditions. These results hold for a generic number of species (see Supplementary Information). For both these panels the kernels are computed explicitly from the network adjacency matrix (see Methods), $e_1 = 0.25$, and $e_2$ is computed via the central manifold condition.}
	\label{fig:centralmanifold}
\end{figure}

Although we can trivially extend the notion of metapopulation capacity for each species $\lambda_M^\alpha$, the condition $\lambda_M^\alpha > e_\alpha$ no longer guarantees species survival. Rather, as we prove in the Supplementary Information, species survival now depends on the ratio $\ev{r_{\alpha}} = \ev{K}_\alpha/e_\alpha$, where $\ev{K}_\alpha = N^{-2} \sum_{ij} K_{\alpha, ij}$. As $\ev{r_{\alpha}}$ quantifies the balance between colonization and extinction, it describes the \emph{average species fitness}. In Figure \ref{fig:centralmanifold}a, we show the phase plot in the $(e_1, e_2)$ space for two species in a grid dispersal network. In this case, the network is highly homogeneous, and the dispersal kernel is invariant under translations. We find that the species with the highest fitness typically survives in the long-time limit, leading to stable monodominance. However, coexistence is possible by fine-tuning the average species fitnesses to be equal (Figure \ref{fig:centralmanifold}a, solid black line). This requires a precise trade-off between dispersal and extinction, i.e., $\ev{K}_\alpha/e_\alpha = \ev{K}_\beta/e_\beta$ for all species pairs $\alpha$, $\beta$. As we explicitly prove in the Supplementary Information, this stationary coexistence state is a central manifold, so the patches where a species survives are solely determined by its initial state (\cref{fig:centralmanifold}b).

However, this fine-tuned coexistence is not possible in less homogeneous dispersal networks. In \cref{fig:centralmanifold}c we show the evolution of two species in an Erd\H{o}s-R\'enyi dispersal network. Even if their average fitness is equal, the ecosystem reaches monodominance after displaying a metastable state in which the two species only temporarily coexist. Although the lifetime of this metastable state increases with network size $N$ (see Supplementary Information), it is always one species that survives and colonizes the whole network at stationarity, independently of the initial state. Hence, coexistence in general landscapes is not feasible when all habitat patches are identical.

\subsection*{Stable coexistence in heterogeneous landscapes}
\noindent To understand how landscape heterogeneity shapes ecosystem diversity, we now turn to the general case in which $e_{\alpha i}$ depends also on the habitat patch $i$. In order to derive analytical insights, we first consider the mean field limit of the model, where all patches are completely connected in a large ecosystem, i.e., $N \to \infty$. In this scenario, the dispersal kernel reads $K_{\alpha, ij} = K_{\alpha}/N$ (see Methods). The stationary state $p^*_{\alpha i}$ obeys the consistency equation
\begin{equation}
\label{eqn:consistency_eq}
    1 = \frac{1}{N} \sum_{i=1}^N r_{\alpha i} \left( 1 + \sum_{\beta=1}^S r_{\beta i} \langle p^*_\beta \rangle \right)^{-1},
\end{equation}
where angular brackets indicate averaging over patches, and $r_{\alpha i} = K_{\alpha}/e_{\alpha i}$ is now the \emph{local species fitness}, which quantifies the balance between colonization and extinction on each patch rather than across all the patches (see Methods). We assume that, for a given species $\alpha$, the values $r_{\alpha i}$ appearing in \cref{eqn:consistency_eq} are extracted from a generic probability distribution $P_r(r | \vec{\zeta}_{\alpha})$, where $\vec\zeta_\alpha$ are the species-dependent parameters on which $P_r$ depends. $P_r$ describes the landscape heterogeneity in terms of habitat-dependent colonization and extinction. Furthermore, with this quenched disorder assumption, we can rewrite \cref{eqn:consistency_eq} as
\begin{equation}
1= S \int_0^\infty  \mathrm{d}z e^{-S \bar{F}(z, \vec{x})} \left( -\frac{W'_{\alpha}(z \, x_{\alpha})}{W_\alpha(z \, x_{\alpha})} \right)
\label{eqn:integral-consistency_eq}
\end{equation}
where $x_\alpha = S \ev{p^*_\alpha}$, $W_\alpha(\omega)$ is the moment generating function of $P_r(r|\vec{\zeta}_{\alpha})$ and $\bar{F}(z, \vec{x}) = z - \frac{1}{S} \sum_{\beta=1}^S \ln W_\beta (z \, x_\beta)$.

\begin{figure*}[t]
    \centering
    \includegraphics[width=\textwidth]{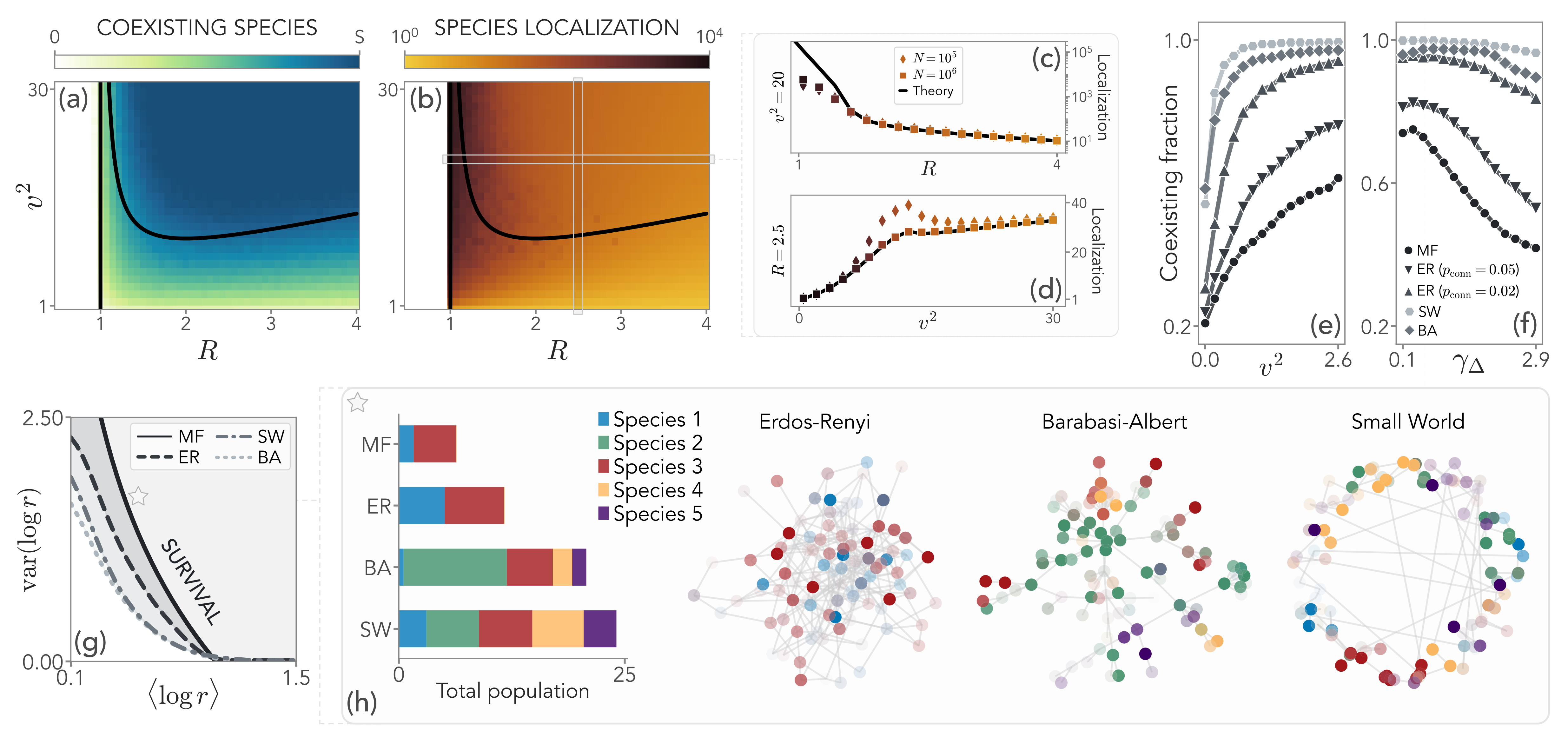}
    \caption{\textbf{Strong spatial heterogeneities lead to stable coexistence through the emergence of ecological niches}. (a) Coexistence in an all-to-all dispersal network with $N = 10^5$ patches and $S = 20$ species, where $R$ is the baseline species fitness and $v^2$ is the variance of the disorder, which follows a log-normal distribution of mean $\ev{r_\alpha} = R + \Delta_\alpha/S$. Coexistence of all species is possible at strong enough disorder, in agreement with theoretical predictions (black lines). Below $R=1$, no species can survive. Here we take $\Delta_\alpha$ to be evenly spaced between $\pm 5/2$, so that $\gamma_\Delta = 2.5$. (b) Coexistence is enabled through species localization (computed via the inverse participation ratio), signaling the spontaneous emergence of ecological niches. (c) Close to the extinction line at $R = 1$, species become strongly localized in a few habitat patches, as predicted by the theory in the $N\to \infty$ limit. (d) For large ecosystems, localization increases with the disorder variance $v^2$. At low enough $N$, localization displays a local peak at intermediate values of the variance due to finite-size effects. (e) The fraction of coexisting species, at constant average fitness, strongly depends on the topology of the dispersal network. Small-world (SW) and Barabasi-Albert (BA) networks allow for the coexistence of more species at lower values of the variance. Here, $R = 2$, $\gamma_\Delta = 1$, $S = 5$, and $N = 100$. (f) As the species heterogeneity $\gamma_\Delta$ increases, it becomes harder to sustain more diverse species and topology plays a fundamental role in determining their coexisting fraction. Parameters are as in the previous panel, with $v^2 = 3$. (g) At $\gamma_\Delta = 0$, the average fitness across species is equal, and we can focus on the effect of topology on the extinction transition. We find that, in general, topology helps the survival of the metacommunity. (h) In particular, Barabasi-Albert and small-world networks display more diversity and a higher total population compared to the mean-field and Erdos-Renyi counterparts. Simulations and parameters of the dynamical model are specified in the Methods.}
    \label{fig:phaseplot-and-loc}
\end{figure*}

For a large ecosystem, \cref{eqn:integral-consistency_eq} can be solved exactly as a $1/S$ expansion for the stationary state. As an illustrative and interesting case, we derive explicit results when $P_r(r | \vec{\zeta}_{\alpha})$ is a log-normal distribution with variance $\sigma^2_r=v^2$ and mean $\ev{r_\alpha}=R + \Delta_\alpha /S$. The variance $v^2$ quantifies the disorder strength, $R$ sets the baseline fitness, and the vector $(\Delta_1, \dots, \Delta_S)$ measures the deviation of each species from the baseline. As we show in the Supplementary Information, when $S$ is sufficiently large, the coexistence of all species becomes possible only if:
\begin{equation}
R > 1 \quad \textrm{and} \quad v^2 > v_c^2 := \gamma_\Delta \, \frac{R^2}{R - 1}.
\label{eqn:critical-v}
\end{equation}
where the parameter $\gamma_\Delta = \sum_{\alpha=1}^S\Delta_\alpha/S - \min_\alpha \Delta_\alpha$ determines the extent to which species are different from one another. The first of the two conditions in \cref{eqn:critical-v} indicates that the baseline fitness must be large enough to allow for coexistence. The second, instead, sets a minimal level of disorder strength, which must exceed a critical value $v_c^2$. The more species are heterogeneous, the more disordered the system needs to be for them to survive. Notably, when the average fitness is identical for all species $\gamma_\Delta = 0$, this implies that $v_c^2 = 0$ and coexistence is guaranteed whenever $R > 1$. In \cref{fig:phaseplot-and-loc}a, we show the number of coexisting species in the $(R, v^2)$ phase space for a given choice of $\Delta_\alpha$, obtained numerically for a finite number of habitat patches and species. The results are in excellent agreement with the theoretical prediction (black lines). In particular, no species survives if $R < 1$, as the baseline fitness is too small to support survival, whereas full coexistence is possible only at large enough disorder strength.

Remarkably, in such disorder-induced coexistence conditions, species densities are not evenly spread throughout the landscape. Rather, the largest share of each species is concentrated within a fraction of the available habitat patches. We measure this effect by computing how localized each species is (see Methods). In \cref{fig:phaseplot-and-loc}a-d, we show two fundamental results. First, as the ecosystem approaches the $R = 1$ line, the species localization drastically increases. This phenomenon is well predicted by the theory and has profound ecological and physical consequences. In fact, the boundary at $R = 1$ marks a sharp transition towards widespread extinction, and, to be able to survive near this extinction threshold, species must maximize their segregation. Second, we observe an increase in species localization as the heterogeneity of the landscape, i.e., the disorder variance $v^2$, becomes larger. Thus, we find that a stronger landscape heterogeneity fosters the spontaneous emergence of ecological niches. These niches are the key to coexistence, allowing species to minimize the detrimental effect of competition for space by specializing in a fraction of the available habitat patches.

\subsection*{Landscape structure promotes coexistence}
\noindent Our analytical results have been obtained for an all-to-all dispersal network, and in the limit of a large spatial ecosystem, i.e., $N \to \infty$. We show that they qualitatively hold even when both $N$ and $S$ are small and in non-trivial dispersal network topologies. We study the impact of real-world network motifs by comparing the mean-field (MF) scenario with three prototypical networks: Erd\H{o}s-R\'enyi (ER) networks at two different connectances, to introduce network sparsity; Barabási-Albert (BA) networks, characterized by hubs; and small-world (SW) networks, with high clustering coefficient \cite{newman2018networks}. To compare the effects of different dispersal topologies, we keep the mean fitness $\ev{r_\alpha}$ for each species constant, which results in similar single-species survivability across different networks (see Methods).

We find that the results of the MF are still qualitatively valid, with the fraction of coexisting species increasing with landscape heterogeneity across all networks (\cref{fig:phaseplot-and-loc}e). However, the dispersal structure quantitatively shifts the coexistence transition. ER networks display enhanced coexistence with increasing sparsity (\cref{fig:phaseplot-and-loc}e), a trend further detailed in the Supplementary Information. At a fixed sparsity, the introduction of hubs or small-world topological features further boosts species' coexistence. Another key result from \cref{eqn:critical-v} is that there is a maximum amount of species diversity, characterized by the heterogeneity parameter $\gamma_\Delta$, that can be sustained at a given strength of landscape disorder. In \cref{fig:phaseplot-and-loc}f, we show that the coexisting fraction of species at fixed $v^2$ rapidly declines when the species fitness heterogeneity increases. Switching to a sparser ER network substantially enhances the coexisting fraction at higher $\gamma_\Delta$, while BA and SW networks are able to support even more diverse species.

To remove the disadvantage due to fitness differences towards coexistence, we also consider the case of neutral species with equal fitness on average, given by $\gamma_\Delta = 0$ (see Methods). In \cref{fig:phaseplot-and-loc}g, we show that changing the topology shifts the extinction transition line, allowing for survival with less spatial disorder. Indeed, for the same disorder realization, structured landscapes such as SW and BA networks support a larger number of species as well as a higher total population (\cref{fig:phaseplot-and-loc}h). 
Interestingly, although all species in this example have the same average fitness, the interplay between spatial disorder and network structure determines which species survive, leading to uneven population distributions. For instance, the second species in \cref{fig:phaseplot-and-loc}h becomes dominant in the BA network because of its higher local fitness within the largest hub.
 
Hence, landscape structure plays a fundamental role in shaping species' survival when dispersal and colonization are both taken into account. Depending on the interplay between local species fitness and dispersal topology, a species may thrive in a given network but go extinct in another. Importantly, the presence of the disorder-induced coexistence transition across different networks shows that our exact results are still valid well beyond the mean-field case.

\subsection*{Spatial patterns in heterogeneous landscapes with correlated disorder}
\noindent Realistic landscapes not only consist of structured dispersal networks but also a spatial correlation in environmental factors. To investigate our results under this constraint, we introduce a correlation between a species' fitness in patch $i$ and patch $j$ which decreases with distance, with a typical correlation length of $d_\mathrm{corr}$ (see Methods). 

\begin{figure}[t]
    \centering
    \includegraphics[width=\columnwidth]{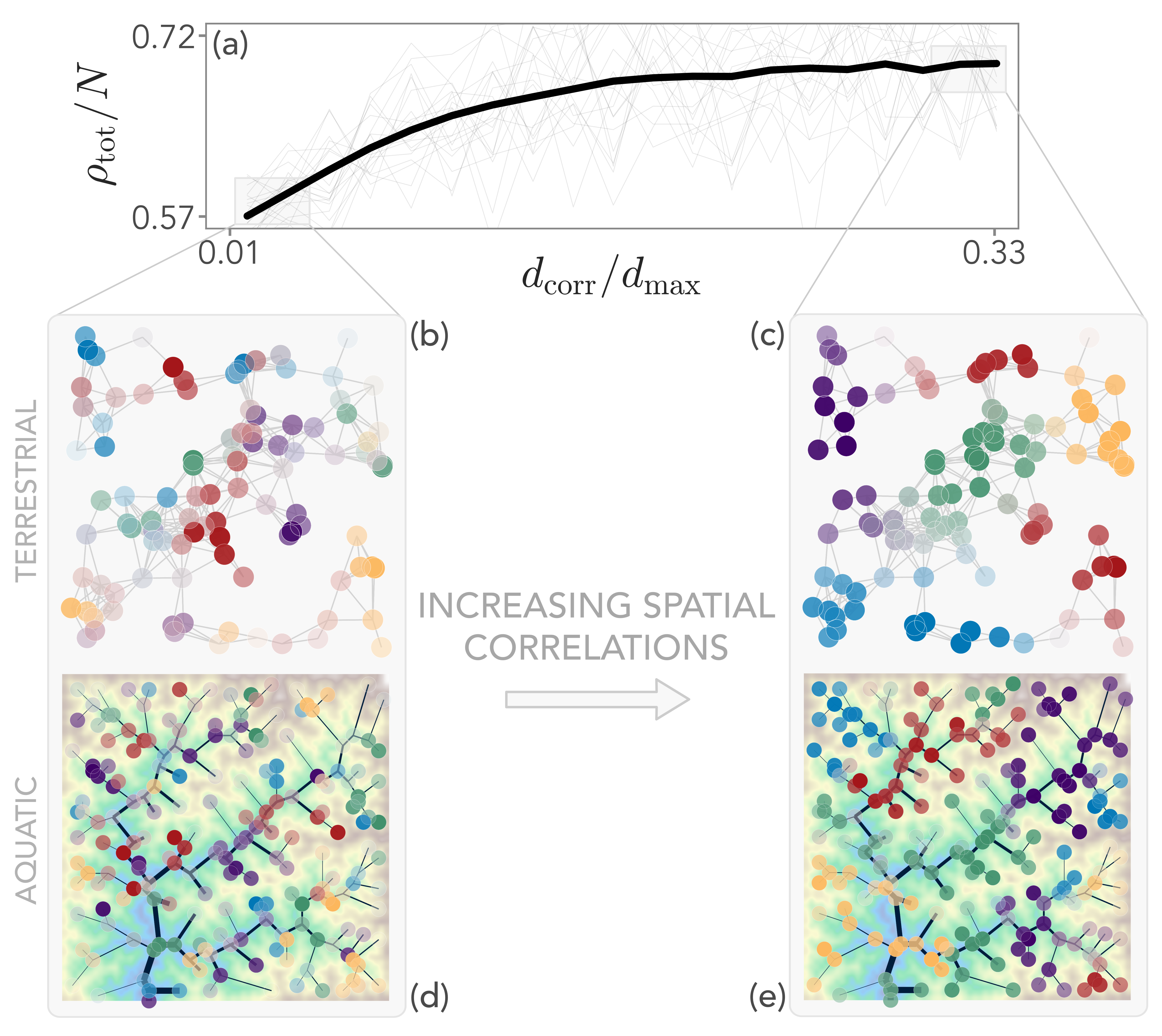}
    \caption{\textbf{Effect of spatially correlated disorder in niche formation}. (a) The total population $p_\mathrm{tot}$ increases with the spatial correlation length of the disorder over the maximum distance between nodes, $d_\mathrm{corr}/d_\mathrm{max}$. (b-c) We first study the effect of spatial correlations on terrestrial landscapes, modeled through a random geometric graph (RGG). The correlation of the disorder decays with the Euclidean distance between the nodes and allows for the spatial clustering of niches. As a result, nodes that are close together in space are occupied by the same species (different node colors represent different species). (d-e) In aquatic landscapes, obtained via optimal channel networks (OCNs), spatial correlations decay instead with the network distance. Now, niches for different species spontaneously emerge in different river branches, since nodes that are spatially close to each other may be at large network distances. The shaded terrain map represents the elevation map obtained from the OCN.}
    \label{fig:spatialnets}
\end{figure}

We consider two kinds of landscape: terrestrial and aquatic. We model terrestrial landscapes using random geometric graphs (RGGs) \cite{dall2002random}. Each habitat patch is embedded in a random spatial position, and the patches are connected if their Euclidean distance is smaller than a given threshold. Furthermore, exploration through these dispersal pathways is inversely proportional to their distance. In \cref{fig:spatialnets}a, we show that, on average, a larger correlation length of the disorder increases the total population in the ecosystem. This increase is fostered by a spatial clustering of niches, such that the same species occupies habitat patches that are close together (\cref{fig:spatialnets}b-c), making colonization more efficient overall and thereby also reducing interspecies competition. 

The features of aquatic landscapes, instead, have been extensively shown to be well-captured by optimal channel networks (OCNs) \cite{rigon1993optimal, rinaldo2020}. In this case, patches represent fluvial habitats and are connected by a river network flowing from high to low elevations (see Methods). We assume that the correlation between species fitness decays with the network distance rather than the Euclidean one, which quantifies the dendritic connections along the river network. In \cref{fig:spatialnets}d-e, we see that again correlations induce the emergence of spatial ordering. This time, nearby spatial regions may be occupied by different species, as spatially close habitat patches may be distant along the river network. Thus, the emergent ecological patterns in space intrinsically reflect the underlying landscape structure.

\section*{Discussion}
\noindent In this work, we studied a general theoretical framework to characterize coexistence in ecosystems dominated by directional dispersal induced by the connectivity of habitat patches. We find that stable coexistence is favored by heterogeneous landscapes endowed with quenched disorder, and it is enabled by the spontaneous emergence of ecological niches that minimize direct competition. The topological and spatial structure of dispersal networks may prompt more coexistence, therefore playing a pivotal role in shaping the features of dendritic ecosystems. Crucially, we find that a sufficient degree of landscape heterogeneity is essential for sustaining biodiversity.

Our results show the varied effects that dispersal can have in dendritic metacommunities. When fine-tuned to be in a trade-off with local extinction in homogeneous networks and environments, dispersal can foster a large degree of diversity of coexisting species. However, even small perturbations from this fine-tuned coexistence lead to dispersal promoting interspecific competition, thereby ultimately leading to monodominance. Interestingly, in our model monodominance arises due to competition for space, which can be seen as a single limiting resource. This is reminiscent of the competitive-exclusion principle, which implies that the number of species cannot exceed the number of resources \cite{hutchinson1961paradox}. Yet, in the presence of spatial heterogeneity, increased average local fitness can still enable coexistence if exposed to sufficiently strong spatial disorder. Indeed, in this case, habitat patches may be viewed as multiple resources arising from landscape heterogeneity, allowing for large-scale coexistence without violating the competitive exclusion principle.

Our results demonstrate that dispersal can act as a stabilizing force in general metacommunities. This applies when the underlying substrate for dispersal and ecological interactions acts synergistically with landscape heterogeneity to reduce interspecific competition \cite{zhang2021dispersal, herberich2023environmental, gibbs2022coexistence}. Although our considerations are limited to spatial competition, they provide a baseline for the impact of dispersal, allowing future works to incorporate it alongside direct ecological interactions among species \cite{luo2022multispecies}. One of the key predictions of our model is that, as the landscape parameters are driven toward widespread extinction, species localization will rapidly increase. This is akin to typical early warning signals of critical transitions in ecological systems \cite{scheffer2012anticipating}, possibly serving as an indicator of the health of the ecosystem.

Finally, we find that the structure of dispersal networks may enable broader niche creation and reduce competition, thereby promoting beneficial coexistence in motifs such as hubs and clustered nodes. Indeed, weak and sparse interactions in local communities have been shown to be beneficial to coexistence \cite{zhang2021dispersal}. In particular, we find that landscapes represented by terrestrial or aquatic networks lead to the emergence of characteristic spatial patterns in correlated environments.

In the last twenty years, it has become clear that dispersal mechanisms can fundamentally alter the structure of ecosystems, both with and without disorder \cite{rinaldo2020}. In this context, our framework allows for deeper insight into quantitative characterizations of individual-level processes underlying colonization and extinction, while being simple enough to allow for analytical treatment. As such, several extensions could be readily considered, allowing for in-depth characterizations of specific ecosystems and the introduction of more complex forms of species interactions beyond competition for space. In particular, mutualistic interactions may reduce the critical value of spatial heterogeneity needed for biodiversity. Our approach will enable us to study how landscape structure affects cooperation and competition between different ecological niches, a pressing matter in understanding how biodiversity evolves under environmental changes.

\begin{acknowledgments}
\noindent G.N. and A.R. acknowledge funding provided by the Swiss National Science Foundation through its Grant CRSII5\_186422. D.B., A.M., and S.A. acknowledge the support of the Italian  Ministry of University and Research (project funded by the European  Union - Next Generation EU: “PNRR Missione 4 Componente 2, “Dalla ricerca all’impresa”, Investimento 1.4, Progetto  CN00000033”). 
S.S acknowledges financial support under the National Recovery and Resilience Plan (NRRP), Mission 4, Component 2, Investment 1.1, Call for tender No. 104 published on 2.2.2022 by the Italian Ministry of University and Research (MUR), funded by the European Union – NextGenerationEU – Project Title: Anchialos: diversity, function, and resilience of Italian coastal aquifers upon global climatic changes – CUP C53D23003420001 Grant Assignment Decree n. 1015 adopted on 07/07/2023 by the Italian Ministry of Ministry of University and Research (MUR).
\end{acknowledgments}

\setcounter{equation}{0}
\makeatletter
\renewcommand{\theequation}{S\arabic{equation}}
\renewcommand{\thefigure}{S\arabic{figure}}
\renewcommand{\thesection}{S\Roman{section}} 
\section*{Methods}

\noindent\textbf{Metacommunity model with dispersal.} We start from an individual-based dynamics describing $S$ species in $N$ interconnected habitat patches. Each species, in principle, may explore the patches differently, according to its own dispersal pathways, so that the overall dispersal network is a multilayer network. For simplicity, here we assume that all species follow the same dispersal network, but our framework can be immediately generalized to other cases. Each patch has a finite number $M$ of colonizable sites. We denote with $P_{\alpha i}$ an individual of the $\alpha$-th species settled in a site of patch $i$. Such individuals give birth to offspring $X_{\alpha i}$ that explore the network of habitat patches before attempting to settle and colonize. The model is described by the reactions
\begin{equation}
    \begin{gathered}
        P_{\alpha i} \xrightarrow{e_{\alpha i}} \varnothing_i, \quad\quad P_{\alpha i} \xrightarrow{c_{\alpha i} h_\alpha A_{ij}} P_{\alpha i} + X_{\alpha j} \\
    X_{\alpha i} \xrightarrow{D_{\alpha} A_{ij}} X_{\alpha j} \\ X_{\alpha i} + \varnothing_i \xrightarrow{\lambda_\alpha / M} P_{\alpha i} , \quad X_{\alpha i} + P_{\alpha i} \xrightarrow{\lambda_\alpha / M} P_{\alpha i}
    \label{reactioneqs}
    \end{gathered}
\end{equation}
where, for species $\alpha$, $e_{\alpha i}$ is the extinction rate in patch $i$, $c_{\alpha i}$ is the colonization rate, $D_\alpha$ and $\lambda_\alpha$ are respectively the exploration and settling rate, $h_\alpha$ is the feasibility of exploration of the species, and $A_{ij}$ is the adjacency matrix of the dispersal network that connects habitat patches.

As in previous works \cite{NicPad2023}, we can obtain an explicit metapopulation model in the limit of fast exploration \cite{nicoletti2024information}. If $p_{\alpha i} = \ev{P_{\alpha i}}/M$, we can derive \cref{eqn:model-general} exactly from the leading order of the Kramers-Moyal expansion of the master equation (see Supplementary Information) \cite{gardiner}. In particular, the kernel is given by the matrix
\begin{equation}
\label{eqn:M:kernel}
    K_{\alpha\beta, ij} = \delta_{\alpha\beta}h_{\alpha} \sum_{l = 1}^N A_{jl} \sum_{k = 1}^N \frac{V_{ik} (V^{-1})_{kl}}{1 + f_\alpha \omega_k}
\end{equation}
where $\omega_k$ is the $k$-th eigenvalue of the transpose out-degree Laplacian of the dispersal network, and $V_{ij}$ is the matrix of its right eigenvectors. The parameter $f_\alpha = D_\alpha /\lambda_\alpha$ represents the exploration efficiency of species $\alpha$ - if $f_\alpha \gg 1$, explorers will visit many habitat patches before attempting colonization, whereas if $f_\alpha \ll 1$ they will remain close to the originating patch. In simulations of the model, we typically set $h_\alpha = \xi_\alpha /(1 + f_\alpha^{-1})$, with $\xi_\alpha$ the maximal dispersal capacity of the species \cite{NicPad2023}. In this way, we ensure that exploration is not possible as $f_\alpha \to 0$, and that $f_\alpha \to \infty$ gives a finite kernel.

\noindent\textbf{Mean-field dispersal kernel.} In a mean-field network, we write the adjacency matrix as $A_{ij} = 1 - \delta_{ij}$ and we rescale $\xi_\alpha \to \xi_\alpha/N$. Then, the kernel elements are given by
\begin{equation*}
    K_{\alpha, ij}^\mathrm{MF} = h_\alpha\left[\frac{(N-1)f_\alpha}{1 + N f_\alpha} \delta_{ij} +  \frac{1 + (N-1)f_\alpha}{1 + N f_\alpha}\left(1 - \delta_{ij}\right) \right]
\end{equation*}
so that, in the large $N$ limit, we find that $K_{\alpha, ij} = K_\alpha / N$ for all edges $i$ and $j$, with $K_\alpha = \xi_\alpha /(1 + f_\alpha^{-1})$.

\noindent\textbf{Fine-tuned coexistence.} We consider a homogeneous landscape, where all habitat patches have the same extinction rate $e_\alpha$ for a given species, and the dispersal network is invariant under translations. In this scenario, the stationary species density cannot explicitly depend on the habitat patches, so $p^*_{\alpha i} = p^*_\alpha$. A solution $p^*_\alpha > 0$ must obey the self-consistency equation
\begin{equation}
    1 - \sum_{\beta = 1}^S p_\beta^* = \frac{e_\alpha}{N \ev{K_\alpha}}, \; \forall \alpha = 1, \dots, S
\end{equation}
where $\ev{K_\alpha} = N^{-2} \sum_j K_{\alpha, ij}$ does not depend on $i$ due to the underlying translational invariance. Thus, coexistence is possible if and only if the average species fitness $\ev{r_\alpha} = \ev{K_\alpha}/e_\alpha$ is identical for all species. In the Supplementary Information, we prove that this stationary solution corresponds to a zero eigenvalue of the Jacobian, and thus is a central manifold. Hence, this comprises a family of stationary solutions that explicitly depend on the initial condition, which disappears in the absence of translational invariance.

\noindent\textbf{General solution in heterogeneous landscapes.} The mean-field equation corresponding to \cref{eqn:model-general} are given by
\begin{equation}
	\dot{p}_{\alpha i}  = e_{\alpha i} \left[ - p_{\alpha i} + \left(1 - \sum_{\beta=1}^{S} p_{\beta i} \right) r_{\alpha i} \ev{p_\alpha}\right],
\end{equation}
where $\ev{p_\alpha} = \sum_{j = 1}^N p_{\alpha j}/N$, and we introduced the local species fitness $r_{\alpha i} = K_\alpha/e_{\alpha i}$. The stationary values $p^*_{\alpha i}$ must clearly obey the consistency equation in \cref{eqn:consistency_eq}. We assume that $r_{\alpha i}$ are quenched random variables extracted from a distribution $P_r(r | \vec{\zeta}_{\alpha})$, where $\vec\zeta_\alpha$ are species-dependent parameters. In the Supplementary Information, we show that the consistency equation can be rewritten in terms of the moment-generating function of the distribution of local fitness
\begin{equation}
	W_{\alpha}(\omega) = \int_0^\infty \mathrm{d}r \, P_r(r | \vec{\zeta}_{\alpha}) \, e^{-r \omega}
\end{equation}
as in \cref{eqn:integral-consistency_eq}. In particular, we take the average species fitness to scale as $\ev{r_\alpha} = R + \Delta_\alpha/S + \mathcal{O}(1/S^2)$, where $R$ is the baseline fitness and $\Delta_\alpha$ represent the deviation from such baseline. Then, the rescaled average stationary population $x_\alpha = S \ev{p_\alpha}$ obeys
\begin{equation}
\label{eqn:M:xalpha}
    x_\alpha = \frac{\Delta_\alpha - H}{v_\alpha^2}R + \mathcal{O}\left(\frac{1}{S}\right)
\end{equation}
where $v_\alpha^2 = \ev{r_\alpha^2}-\ev{r_\alpha}^2$, and
\begin{equation}
    H = \left(\;\overline{\frac{1}{v^2}}\;\right)^{-1}\left[ \left(\,\overline{\frac{\Delta}{v^2}}\,\right) - \frac{R-1}{R^2}\right]
\end{equation}
with $\overline{y} = S^{-1}\sum_\beta y_\beta$ denotes the average over disorder (see Supplementary Information). Thus, from \cref{eqn:M:xalpha}, we have that coexistence is possible if $\Delta_\alpha > H$, which reduces to the set of conditions in \cref{eqn:critical-v} if we take $v_\alpha^2 = v^2$ for all species $\alpha$. In general, we immediately see that at large disorder variance, it is easier to satisfy the coexistence condition. All plots in the mean-field case are obtained by explicitly solving the consistency equation.

\noindent\textbf{Species localization and ecological niches.} We compute species localization through the inverse participation ratio (IPR), defined as
\begin{equation}
    N I_{\alpha} = \frac{\ev{p_\alpha^4}}{\ev{p_\alpha^2}^2} \;.
\end{equation}
If a species is present prevalently in $k < N$ habitat patches, then it is easy to show that the IPR is $N I_\alpha \approx N/k$. Hence, if a species is not localized, we expect $NI_\alpha \approx 1$, whereas if it is present in only one habitat patch we have $NI_\alpha \approx N$. Therefore, the IPR is a measure of localization and thus of how much species survival relies on the emergence of ecological niches. In \cref{fig:phaseplot-and-loc}b, we plot the average $S^{-1} N \sum_\alpha I_\alpha$. In the Supplementary Information, we show that the IPR can be computed exactly as 
\begin{equation}
N I_\alpha = \frac{1}{6} \frac{\int_0^\infty dz \, z^{3} \, e^{-S \bar{F}(z, x)} W^{(4)}(z x_{\alpha})/W(z x_{\alpha})}{\left[\int_0^\infty dz \, z \, e^{-S \bar{F}(z, x)} W^{(2)}(z x_{\alpha})/W(z x_{\alpha})\right]^2}
\end{equation}
where $W^{(m)}$ is the $m$-th derivative of the moment-generating function, and $\bar{F}$ has been defined in the main text.

\noindent\textbf{Dynamics in arbitrary dispersal networks.} In an all-to-all dispersal network, the kernel does not depend on the habitat patches and the system solely depends on the average species fitness. However, this is not true in general, as the dispersal kernel in \cref{eqn:M:kernel} has been shown to depend on all possible paths between pairs of patches \cite{NicPad2023}. In this scenario, the local species fitness is $r_{\alpha i} = N \ev{K_\alpha}/e_{\alpha i}$, where the $N$ prefactor comes from the rescaling $\xi_\alpha \to \xi_\alpha/N$. This definition immediately reduces to the mean-field case when we consider an all-to-all network, and once more can be interpreted as a local balance between colonization and extinction.

To integrate numerically the dynamics in an arbitrary dispersal network, which depends explicitly on the extinction rates, we consider first a quenched realization of the disordered local species fitnesses $r_{\alpha i}$. Then, for each species $\alpha$, we compute the kernel elements $K_{\alpha,ij}$ and its average $\ev{K_{\alpha}} = N^{-2} \sum_{ij}K_{\alpha,ij}$, from which we can get the extinction rates as $e_{\alpha i} = N \ev{K_\alpha}/r_{\alpha i}$. Notice that this explicitly shows that, in order to maintain the average species fitness $\ev{r_\alpha}$ constant, the extinction rates must be tuned in response to the specific kernel, i.e., to the topology of the dispersal network. In particular, in \cref{fig:phaseplot-and-loc}h, we take the parameters of the log-normal distribution for the local species fitness to be $\ev{r}=R=1$, $v^2 = 1.5$. The kernel for each species is computed as in \cref{eqn:M:kernel}, with $\xi_\alpha = 1/N$ and $f_\alpha$ to be uniformly spaced in $[0.5, 2]$. To highlight the effect of the network structure, the disorder realization is kept fixed across the different topologies. 

\noindent\textbf{Terrestrial and aquatic landscapes.} To model realistic terrestrial landscapes, we consider random geometric graphs (RGGs) \cite{dall2002random}. RGGs are generated by sampling $N$ patches uniformly in the unit square, i.e., each patch has a spatial position $(x_i, y_i) \in [0,1]\times[0,1]$. Two patches are connected if their Euclidean distance $d_{ij}$ is smaller than a given threshold $d_\mathrm{th}$. We set $d_\mathrm{th} = 0.17$ for \cref{fig:spatialnets}, but our results are qualitatively independent of this choice, provided that the network is connected and not dense. In this spatially embedded network, we take the weights of each edge to be functions of the distance between the patches, i.e., we write the adjacency matrix as
\begin{equation}
    A_{ij} = \frac{d_\mathrm{min}}{d_{ij}} \in [0,1]
\end{equation}
where $d_\mathrm{min}$ is the minimum distance between two patches of the network. In this way, the exploration rate in \cref{reactioneqs} decreases for habitats that are far apart.

Aquatic and riverine landscapes, instead, are well-modeled by optimal channel networks (OCNs) \cite{rigon1993optimal, rinaldo2020}. An OCN is a spanning tree where each node is associated with an elevation $h_i$ and a drainage area $\mathcal{A}_i$, related to one another by the scaling relation $|\nabla h_i| \propto \mathcal{A}_i^{\gamma - 1}$, with the scaling exponent $\gamma = 1/2$. For a given adjacency matrix $A_{ij}$, the areas are given by $\mathcal{A}_i = \sum_j W_{ji}\mathcal{A}_j + 1$. An OCN then is a spanning tree that minimizes the dissipated energy functional $E_{\rm OCN} = \sum_i \mathcal{A}_i^\gamma$ \cite{rinaldo2020}. In particular, in \cref{fig:spatialnets}, the OCN has been aggregated so that each pixel represents either a source, an outlet, or a confluence \cite{carraro2020generation}. In the case of OCNs, the relevant distance that we use to build the metacommunity model is not the Euclidean distance, as for RGGs, but the network distance.

\noindent\textbf{Spatially correlated disorder.} To study the case of spatially correlated disorder, we start from distance matrix $d_{ij}$ - either Euclidean distance for RGGs or network distance for OCNs. Then, we parametrize the covariance $\Sigma_{ij}$ between two habitat patches $i$ and $j$ as
\begin{equation}
    \Sigma_{ij} = \left[1 - \left(\frac{m_1 d_{ij}}{m_2}\right)^2\right] \exp\left[-\frac{\left(m_1 d_{ij}\right)^2}{2 m_3^2}\right]
\end{equation}
which is a Ricker wavelet and allows for both local correlations and long-range anticorrelations. Then, the local species fitness is distributed as a multivariate log-normal distribution, i.e., $r_{\alpha i} = e^{y_{\alpha i}}$ with $y_{\alpha i}$ a multivariate Gaussian variable $y_{\alpha i} \sim \mathcal{N}(\vec{\mu}, \hat{\Sigma})$. For \cref{fig:spatialnets}, we take $\mu_i = 0.5$ and set $m_1 = 3$, $m_2 = 1.3$, and $m_3 = 1$ for RGGs - resulting in small anticorrelations at long distances - and $m_1 = 0.8$, $m_2 = 5$, and $m_3 = 1$ for OCNs - resulting in positive and exponentially-decaying correlations with the network distance. The emergence of spatial patterns is qualitatively independent of these choices, with the only constraint being that $\hat\Sigma$ must be a semipositive-definite matrix. In \cref{fig:spatialnets}, we take the mean of the multidimensional log-normal distribution to be $\mu_i = 0.5$ for all patches. The simulation of the dynamics for each network is performed with the parameters described above.

\end{document}